\documentclass[sigconf]{acmart}
\usepackage{amsmath}
\usepackage{enumitem}
\usepackage{multirow}
\usepackage{arydshln}
\usepackage{amsmath}
\usepackage{algorithmicx}
\usepackage{algorithm}
\usepackage{algpseudocode}
\usepackage{graphicx} 
\usepackage[utf8]{inputenc} % allow utf-8 input
\usepackage[T1]{fontenc}    % use 8-bit T1 fonts
\usepackage{hyperref}       % hyperlinks
\usepackage{url}            % simple URL typesetting
\usepackage{booktabs}       % professional-quality tables
\usepackage{amsfonts}       % blackboard math symbols
\usepackage{nicefrac}       % compact symbols for 1/2, etc.
\usepackage{microtype}      % microtypography
\usepackage{xcolor}         % colors
\usepackage{bm}
\usepackage{lipsum}
\usepackage{balance}
\AtBeginDocument{%
  }

\copyrightyear{2025}
\acmYear{2025}
\setcopyright{acmlicensed}\acmConference[MM '25]{Proceedings of the 33rd ACM International Conference on Multimedia}{October 27--31, 2025}{Dublin, Ireland}
\acmBooktitle{Proceedings of the 33rd ACM International Conference on Multimedia (MM '25), October 27--31, 2025, Dublin, Ireland}
\acmDOI{10.1145/3746027.3758233}
\acmISBN{979-8-4007-2035-2/2025/10}
\begin{document}

\newcommand{\x}{\mathbf{x}}
\newcommand{\M}{\mathbf{M}}
\definecolor{Magenta}{RGB}{237,2,140}
\definecolor{iccvblue}{rgb}{0.21,0.49,0.74}
\def \projecturl{https://more-med.github.io/}
\def \dataseturl{https://huggingface.co/datasets/WSKINGDOM/MORE}

\title{MORE: \underline{M}ulti-\underline{O}rgan Medical Image \underline{RE}construction
Dataset}

\author{Shaokai Wu}

\email{shaokai.wu@sjtu.edu.cn}
\orcid{0009-0004-0531-1120}
\affiliation{%
  \institution{Shanghai Jiao Tong University}
  \city{Shanghai}
  \country{China}
}

\author{Yapan Guo}
\authornotemark[1]
\orcid{0009-0008-5644-4889}
\email{guo\_yapan@outlook.com}
\affiliation{
  \institution{Suzhou Xiangcheng People’s Hospital}
  \city{Suzhou}
  \country{China}}

\author{Yanbiao Ji}
\orcid{0009-0006-8800-9561}
\email{jiyanbiao@sjtu.edu.cn}
\affiliation{
  \institution{Shanghai Jiao Tong University}
  \city{Shanghai}
  \country{China}}

\author{Jing Tong}
\orcid{0009-0002-9306-163X}
\email{tj_19_hf@sjtu.edu.cn}
\affiliation{
  \institution{Shanghai Jiao Tong University}
  \city{Shanghai}
  \country{China}}

\author{Yuxiang Lu}
\orcid{0009-0002-6344-3880}
\email{luyuxiang_2018@sjtu.edu.cn}
\affiliation{
  \institution{Shanghai Jiao Tong University}
  \city{Shanghai}
  \country{China}}

\author{Mei Li}
\orcid{0009-0006-1039-5693}
\email{mei-li@sjtu.edu.cn}
\affiliation{
  \institution{Shanghai Jiao Tong University}
  \city{Shanghai}
  \country{China}}

\author{Suizhi Huang}
\orcid{0000-0003-0172-6711}
\email{huangsuizhi@sjtu.edu.cn}
\affiliation{
  \institution{Shanghai Jiao Tong University}
  \city{Shanghai}
  \country{China}}

\author{Yue Ding}
\orcid{0000-0002-2911-1244}
\email{dingyue@sjtu.edu.cn}
\affiliation{
  \institution{Shanghai Jiao Tong University}
  \city{Shanghai}
  \country{China}}

\author{Hongtao Lu}
\authornote{Corresponding Authors.}
\orcid{0000-0003-2300-3039}
\email{htlu@sjtu.edu.cn}
\affiliation{
  \institution{Shanghai Jiao Tong University}
  \city{Shanghai}
  \country{China}}

\renewcommand{\shortauthors}{Shaokai Wu et al.}



\begin{abstract}
    CT reconstruction provides radiologists with images for diagnosis and treatment, yet current deep learning methods are typically limited to specific anatomies and datasets, hindering generalization ability to unseen anatomies and lesions. To address this, we introduce the \textbf{M}ulti-\textbf{O}rgan medical image \textbf{RE}construction (\textbf{MORE}) dataset, comprising CT scans across 9 diverse anatomies with 15 lesion types. This dataset serves two key purposes: (1) enabling robust training of deep learning models on extensive, heterogeneous data, and (2) facilitating rigorous evaluation of model generalization for CT reconstruction. We further establish a strong baseline solution that outperforms prior approaches under these challenging conditions.
    Our results demonstrate that: (1) a comprehensive dataset helps improve the generalization capability of models, and (2) optimization-based methods offer enhanced robustness for unseen anatomies. The MORE dataset is freely accessible under CC-BY-NC 4.0 at our project page \textcolor{Magenta}{\projecturl}.
\end{abstract}


\begin{teaserfigure}
  \setlength{\abovecaptionskip}{1pt}
  \includegraphics[width=\textwidth]{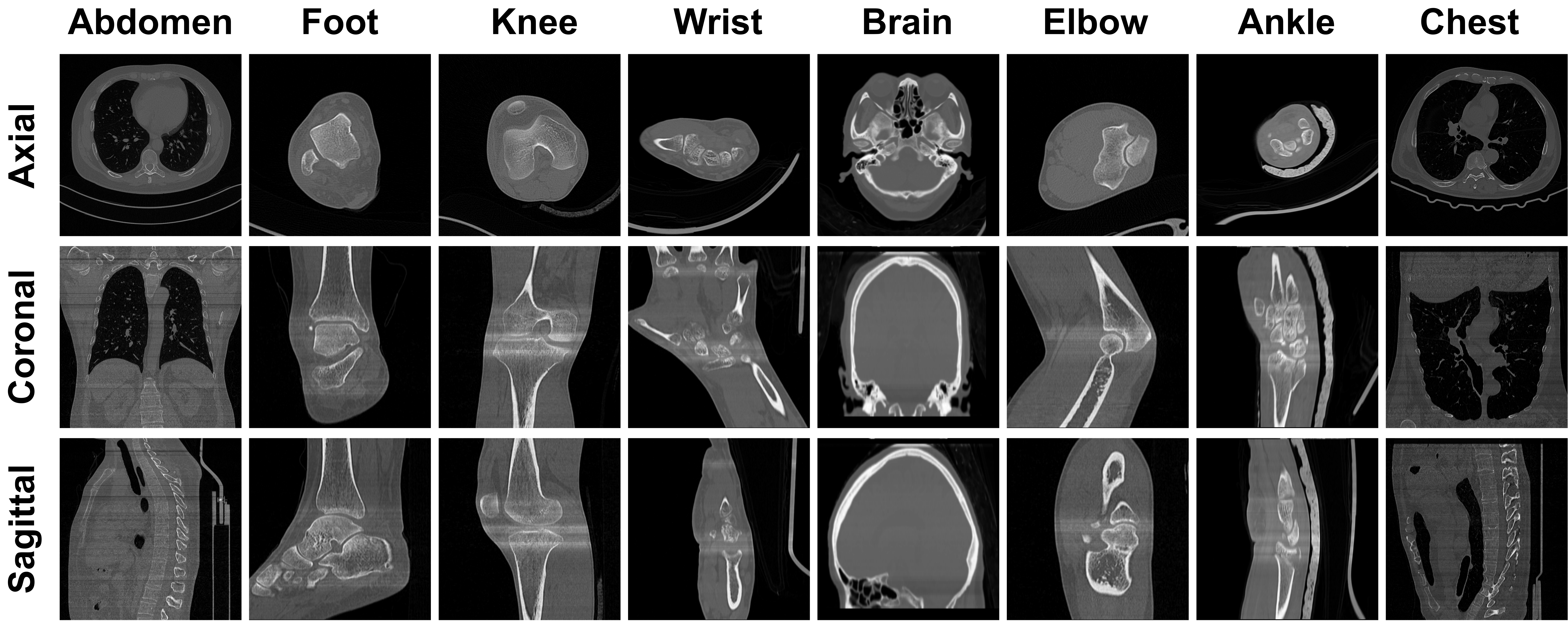}
  \caption{We present CT examples of different anatomical regions from our MORE dataset, with the three rows displaying visualization results in the axial, coronal, and sagittal planes, respectively.}
  \label{fig:teaser}
\end{teaserfigure}

\maketitle

\vspace{-3pt} \section{Introduction}
Computed Tomography (CT)~\cite{nature-ct,nature-ct2} is a cornerstone of modern medical diagnosis, enabling detailed cross-sectional imaging of internal structures through X-ray attenuation measurements~\cite{warren1990x, puttagunta2021medical, zhang2020review, illposed}. Traditional reconstruction methods—such as Filtered Back Projection (FBP)~\cite{fbp1967,fbp2001principles,fbp2019} and Iterative Reconstruction (IR)~\cite{CTreview}—require dense measurements that impose high radiation risks. In contrast, deep learning techniques~\cite{HDNet,UNetct,wavelet1,GMSD} demonstrate significant potential for reconstructing high-quality images from sparse measurements~\cite{CT-Survey-2023, SWORD, DiffusionMBIR}.

However, the efficacy of AI-driven CT reconstruction is critically hampered by a fundamental limitation: the scarcity of comprehensive, clinically diverse datasets. As shown in Table~\ref{dataset}, existing public datasets (e.g., AAPM-Mayo LDCT~\cite{mayo}) are typically restricted to narrow anatomical coverage (e.g., chest/abdomen) and limited pathological representation. This severely constrains model generalization, while clinical practice necessitates robustness across variable anatomies (e.g., limb, lung, liver) and unforeseen pathologies. When models trained on such data encounter new anatomical structures or lesions—common in real-world scenarios—their performance is compromised significantly.

This dataset scarcity impedes progress in two key ways. The first is Generalization Failure. Models overfit to specific anatomies or pathologies, struggling with out-of-distribution cases~\cite{DiffusionMBIR}. For example, models trained exclusively on chest CT scans typically generalize poorly to head scans, and models trained with normal patient samples can perform poorly for those with diseases or lesions. These out-of-distribution (OOD) cases are prevalent and pose a barrier to the generalization ability of deep learning models. The second is the Incomplete Evaluation problem. Current benchmarks cannot assess robustness across clinically relevant variations in age, sex, anatomy, or pathology, and it is hard to justify whether a model is really excellent without a comprehensive evaluation.

\begin{figure}[t]
    \centering
     \setlength{\abovecaptionskip}{5pt}
     \setlength{\belowcaptionskip}{-20pt}
    \includegraphics[width=0.99\linewidth]{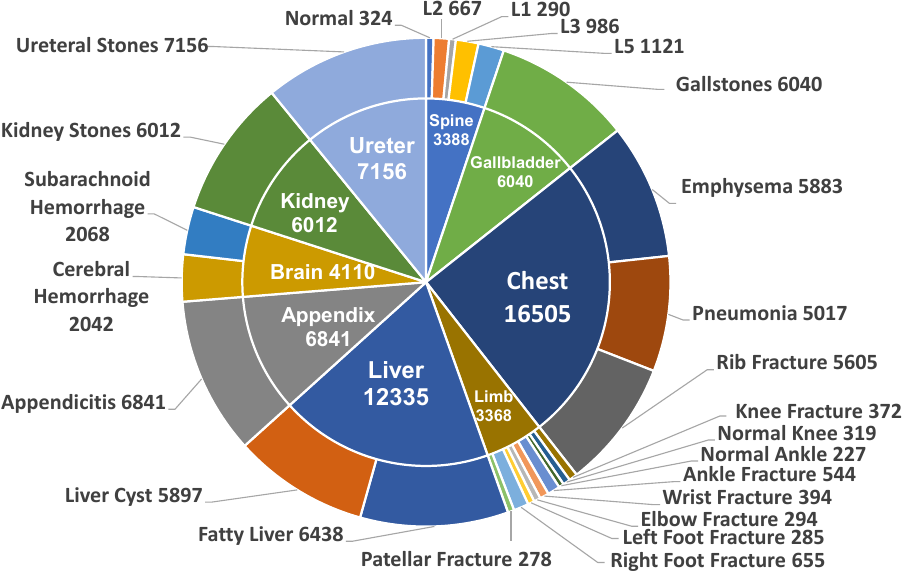}
    \caption{Data distribution of the \textbf{MORE} CT part, containing 9 anatomies and 15 lesion types.}
    \label{fig:ct-diversity}
\end{figure}

Motivated by this, we introduce the \textbf{M}ulti-\textbf{O}rgan tomographic \textbf{RE}construction \textbf{(MORE)} dataset, a comprehensive CT reconstruction dataset designed explicitly for robust, multi-anatomical reconstruction as shown in Figure~\ref{fig:ct-diversity}. It has the following characteristics:
\begin{itemize}[leftmargin=*,topsep=0pt]
    \item Diverse anatomical regions (Chest, Limb, Brain, etc.);
    \item Distinct lesion types (Emphysema, cysts, calcifications, etc.);
    \item Clinical diversity in age, sex, and disease manifestation.
\end{itemize}

Our MORE dataset bridges the critical gap between narrowly focused CT resources and the holistic needs of clinical AI deployment. It not only empowers the training of AI models capable of handling diverse anatomies and lesions but also enables the rigorous benchmarking of reconstruction methods across varied clinical scenarios.

To demonstrate MORE's capabilities, we benchmark representative methods spanning diverse paradigms: traditional Filtered Back Projection (FBP)~\cite{fbp1967}, CNN-based approaches like RED-CNN~\cite{REDCNN}, diffusion-based methods (MCG~\cite{MCG}, DiffusionMBIR~\cite{DiffusionMBIR}, SWORD~\cite{SWORD}), the NeRF-based NeRP~\cite{NeRP}, and the advanced 3D Gaussian Splatting (3DGS)-based method R$^2$-Gaussian~\cite{r2-gaussian}. To further advance reconstruction on this diverse dataset, we propose GIFT (Gaussian Iterative Framework for Tomography), a novel framework that achieves superior performance without pretraining. For reproducibility, our dataset and code are available on our \textcolor{Magenta}{\href{\projecturl}{Project Page}}. 

Our main contributions can be summarized as follows:

\begin{itemize}[leftmargin=*,topsep=0pt]
    \item We introduce a comprehensive multi-organ dataset, \textbf{MORE}, which contains 9 types of anatomies from CT scans with 15 different types of lesions. To the best of our knowledge, it has the most diverse categories compared to existing publicly available datasets. This retrospective study was approved by the Ethics Committee of Suzhou Xiangcheng People's Hospital, China.
    \item We propose a strong baseline solution for CT reconstruction under diverse anatomies and lesions. Our proposed method can iteratively refine the 3D volume with the supervision of the current patient's measurements, possessing strong generalization ability and robustness to unseen anatomies and lesions.
    \item We extensively benchmark various types of methods on the MORE dataset to evaluate their performance. The results show that optimization-based methods outperform deep learning-based methods in terms of generalization ability and robustness to unseen anatomies and lesions.
\end{itemize}

\vspace{-3pt} \section{Related Work}

\noindent \textbf{Task of CT Reconstruction}
Clinical CT imaging captures X-ray projections from multiple angles to reconstruct a 3D volume of internal structures. We simplify the formulation to 2D. The imaging process for a single slice is modeled as:
\begin{equation}
y = Ax + \beta,
\end{equation}
where:

    $x \in \mathbb{R}^{w \times h}$ is the 2D image (width $w$, height $h$),
    $y \in \mathbb{R}^{m \times n}$ is the sinogram ($m$ projection views, $n$ detectors per view),
    $A$ is the Radon transform~\cite{helgason1999radon, he2020radon} matrix, encoding the geometric relationship between the object and the X-ray system,
    $\beta$ represents measurement noise.

The goal is to estimate $x$ from $y$. While the Inverse Radon Transform~\cite{holschneider1991inverse} provides a theoretical solution, noise and underdetermination complicate practical reconstruction. This is typically framed as an optimization problem:
\begin{equation}
x^* = \arg\min_x \mathcal{E}(Ax, y) + R(x),
\end{equation}

\begin{table}[t]
\footnotesize
\setlength{\tabcolsep}{0.7mm}
\caption{Comparison of CT reconstruction datasets (Original: researcher-collected; Derived: from existing databases)}
\label{dataset}
\begin{tabular}{@{}lclc@{}}
\toprule
Dataset & Anatomy & Image Slices & Data Source Type \\ \midrule
MORE & \begin{tabular}[c]{@{}l@{}} Spine, Limb, Gallbladder,\\ Chest, Liver, Appendix,\\ Brain, Kidney, Ureter\end{tabular} & 65,755 & Original \\
AAPM-Mayo LDCT\cite{mayo} &  chest, abdomen, head & 25,141  & Original\\
LoDoPaB-CT~\cite{lodopab-ct} & chest  & 46,573  & Derived \\
Covidx-CT~\cite{Covidnet-ct} & chest  &  104,009  & Derived \\
LIDC/IDRI~\cite{armato2011lung} & chest & 1,018  & Original\\
FUMPE~\cite{masoudi2018new} & chest & 8,792  & Original\\
% JSRT~\cite{shiraishi2000development} & chest & 247  & Original\\
\bottomrule
\end{tabular}
\vspace{-10pt}
\end{table}
% \vspace{-15pt}

\noindent\textbf{CT Reconstruction Datasets}
The availability of large, diverse datasets spanning multiple anatomies and lesion types is fundamental for advancing medical image reconstruction, as it enables robust generalization of AI models across varied clinical scenarios. Established datasets including the AAPM-Mayo LDCT Challenge~\cite{mayo}, LoDoPaB-CT~\cite{lodopab-ct}, COVIDx-CT~\cite{Covidnet-ct}, LIDC/IDRI~\cite{armato2011lung}, and FUMPE~\cite{masoudi2018new} have served as critical benchmarks for methodological development. Nevertheless, these resources exhibit pronounced limitations in anatomical and pathological scope, as quantified in Table~\ref{dataset}, which reveals a concentration on thoracic imaging and sparse representation of abdominal, neurological, or musculoskeletal structures. Similarly, lesion diversity remains restricted primarily to nodules, tumors, and infectious findings, omitting rarer pathologies. This narrow coverage impedes the validation of reconstruction algorithms under realistic, heterogeneous clinical conditions and hinders the development of comprehensive, clinically relevant benchmarks. 
To address these critical gaps, we introduce our Multi-Organ medical image REconstruction dataset, curated to encompass 9 anatomically distinct anatomies and 15 clinically significant lesion types, broadening the scope for model training and evaluation.

\section{Multi-Organ CT REconstruction Dataset}
We organize this section into three distinct parts: First, the background and characteristics of the \textbf{MORE} dataset are introduced. Second, data acquisition and processing are discussed. Then, data usage, access, and ethics and privacy considerations are discussed, ensuring both effective utility and stringent preservation of patient privacy. Finally, we present the benchmark tasks and methods, which serve as a foundation for evaluating the performance of reconstruction methods on this dataset.

\subsection{Background and Characteristics}
\label{subsec:background and characteristics}
This study is a retrospective study, encompassing imaging data from patients who underwent CT scans at the Suzhou Xiangcheng People’s Hospital due to various diseases. This research involved a total of 135 patients. The number of CT image slices is 65,755.
The dataset covers a wide range of anatomies and lesions, including Spine, Knee, Ankle, Wrist, Elbow, Foot, Patella, Gallbladder, Lung, Liver, Appendix, Brain, Kidney, Ureter, and Arachnoid Membrane.

\noindent\textbf{Motivation} The primary motivation behind the creation of the MORE dataset was to address the limitations of existing datasets in the field of medical image reconstruction. Existing datasets are often limited in scope and thus not large enough to train deep learning models effectively. Besides, they cover only a few anatomies or conditions and are thus not comprehensive enough to represent the real-world distribution of medical images. The key features of this dataset are its size, diversity (various anatomies), and comprehensiveness (various lesions). The comparison of various medical image reconstruction datasets is shown in Table~\ref{dataset}.

\noindent\textbf{Patient Demographics} This research involves a total of 135 patients. The median age of the participants was 52 years, ranging from 7 to 85 years. The age distribution is as follows: 0-20 years (5.4\%), 21-40 years (29.5\%), 41-60 years (37.2\%), 61-80 years (24.0\%), and 81-100 years (3.9\%). The gender distribution was 59.7\% male and 40.3\% female.

\noindent\textbf{Data Distribution} Our dataset covers a wide range of anatomies and lesions, including Spine, Knee, Ankle, Wrist, Elbow, Foot, Patella, Gallbladder, Lung, Liver, Appendix, Brain, Kidney, Ureter, and Arachnoid Membrane for CT scans. We show the specific distribution in Figure~\ref{fig:ct-diversity} and some typical samples in Figure~\ref{fig:teaser}.

\noindent\textbf{Multiple Lesions} Our dataset includes a wide range of lesions, such as rib fractures, appendicitis, pneumonia, subarachnoid hemorrhage, emphysema, ureteral stones, kidney stones, cerebral hemorrhage, fatty liver, fractures of the knee, ankle, wrist, elbow, and foot, and spinal fractures. They are distributed across various anatomies, ensuring a comprehensive representation of medical conditions.
\vspace{-5pt}
\subsection{Data Acquisition and Processing}
\noindent\textbf{Staff Configuration} All CT scans were collected and evaluated by three experienced radiologists. The radiologists were responsible for reviewing the scans and identifying any abnormalities or lesions. Among the three radiologists, two were senior radiologists with over 10 years of experience, and one was a junior radiologist with 5 years of experience. The radiologists worked together to ensure the accuracy and consistency of the data.

\noindent\textbf{Data Selection} The CT scans were selected based on the following criteria: (1) the scans were of high quality, with minimal artifacts or noise, (2) the scans covered a wide range of anatomies and conditions, and (3) the scans were representative of the clinical cases encountered in practice. 
Our radiologists first categorized the scans based on the body part imaged and the condition depicted, and then selected typical cases from the corresponding parts, including internal and external medicine and acute and chronic cases.

\noindent\textbf{Scan Parameters} For each individual sample, the window width and window position that are commonly displayed for the corresponding disease type are selected. Samples of two slice thicknesses (1mm and 3mm) are chosen for CT scans. 
The CT scans were acquired using a Siemens SOMATOM Definition AS+ scanner. The detailed scan parameters are included in our \textcolor{Magenta}{\href{https://github.com/MORE-Med/MORE-Med.github.io/blob/main/Assets/ScanParameters.csv}{Scan Parameters}}.

\noindent\textbf{Data processing} 
The image data is provided and is easy to use. Slices within the same sequence can be identified by their file names, and each slice is stored as a 2D array of pixel intensities without extra transformation. Intensity values depend on the type of scan and the scanning parameters. The pixel intensities represent Hounsfield units. To facilitate other researchers' use, we also provide PNG images for each DICOM file which can be easily visualized.

We preprocess the raw DICOM files into 2D image slices following the practice in SimpleITK. These image slices are used as the ground truth, i.e., the target volume for the reconstruction task.

\vspace{-5pt} \subsection{Data Usage, Access, and Ethics}
\label{subsec:data usage access and ethics}
\noindent\textbf{Ethics and Privacy} This study was approved by the ethics committee of Suzhou Xiangcheng People’s Hospital. The approval number is 2024-KY-03. 
Moreover, all DICOM data have been anonymized by the RSNA clinical trial processor to protect patient privacy and comply with the Helsinki Declaration.

\noindent\textbf{Data Usage} The sparse measurements are used as the input for the reconstruction methods. The reconstructed images are compared with the ground-truth images to evaluate the performance of the reconstruction methods. The reconstructed images are evaluated using standard metrics such as Peak Signal-to-Noise Ratio (PSNR)~\cite{psnr} and Structural Similarity Index (SSIM)~\cite{ssim}. 

\noindent\textbf{License and Data Access} The license of the \textbf{MORE} dataset is CC-BY-NC 4.0. The dataset is freely available for non-commercial use and can be downloaded from \url{\dataseturl}. The dataset is provided in the form of DICOM files, which can be easily read and processed using standard medical imaging libraries.

\vspace{-5pt} 
\subsection{Benchmark}
\label{sec:benchmark}
\noindent\textbf{Benchmark Task} Our MORE dataset currently provides 4 benchmark tasks for four different settings of sparse-view CT reconstruction following widely used practice~\cite{PLA, SWORD, DiffusionMBIR}, namely 60-view, 90-view, 120-view, and 180-view sparse measurements to reconstruct the target volume. The corresponding measurements are provided in our dataset. The evaluation metrics are the Peak Signal-to-Noise Ratio (PSNR)~\cite{psnr} and the Structural Similarity Index (SSIM)~\cite{ssim}. 

We split the dataset, allocating 58,528 CT scans to the training \& evaluation set and 7,498 scans to the test set. The test set is composed of 15 different lesions as described in Section~\ref{subsec:background and characteristics}. 

\noindent \textbf{Benchmark Methods} Our evaluated methods show an evolution of this field, from the traditional FBP method~\cite{fbp1967} that is widely used in clinical practice, to the early deep learning-based method REDCNN~\cite{REDCNN}, followed by the more advanced methods DiffusionMBIR~\cite{DiffusionMBIR}, MCG~\cite{MCG}, and SWORD~\cite{SWORD}, and then a NeRF-based method, NeRP~\cite{NeRP}, that implicitly learns the prior from data, followed by the state-of-the-art method R$^2$-Gaussian~\cite{r2-gaussian}, and finally our baseline solution GIFT, which is elaborated in Section~\ref{sec:Gaussian}. Among these methods, REDCNN~\cite{REDCNN}, MCG~\cite{MCG}, DiffusionMBIR~\cite{DiffusionMBIR}, and SWORD~\cite{SWORD} are training-based and need pretraining to learn prior knowledge. On the other hand, FBP~\cite{fbp1967}, NeRP~\cite{NeRP}, R$^2$-Gaussian~\cite{r2-gaussian}, and our proposed method below do not involve any pretraining process. 

\label{ethics}

\vspace{-3pt} 
\section{GIFT: Baseline Solution}
\label{sec:Gaussian}
\begin{figure}[h]
    \centering
    \setlength{\abovecaptionskip}{0pt}
    \includegraphics[width=0.99\linewidth]{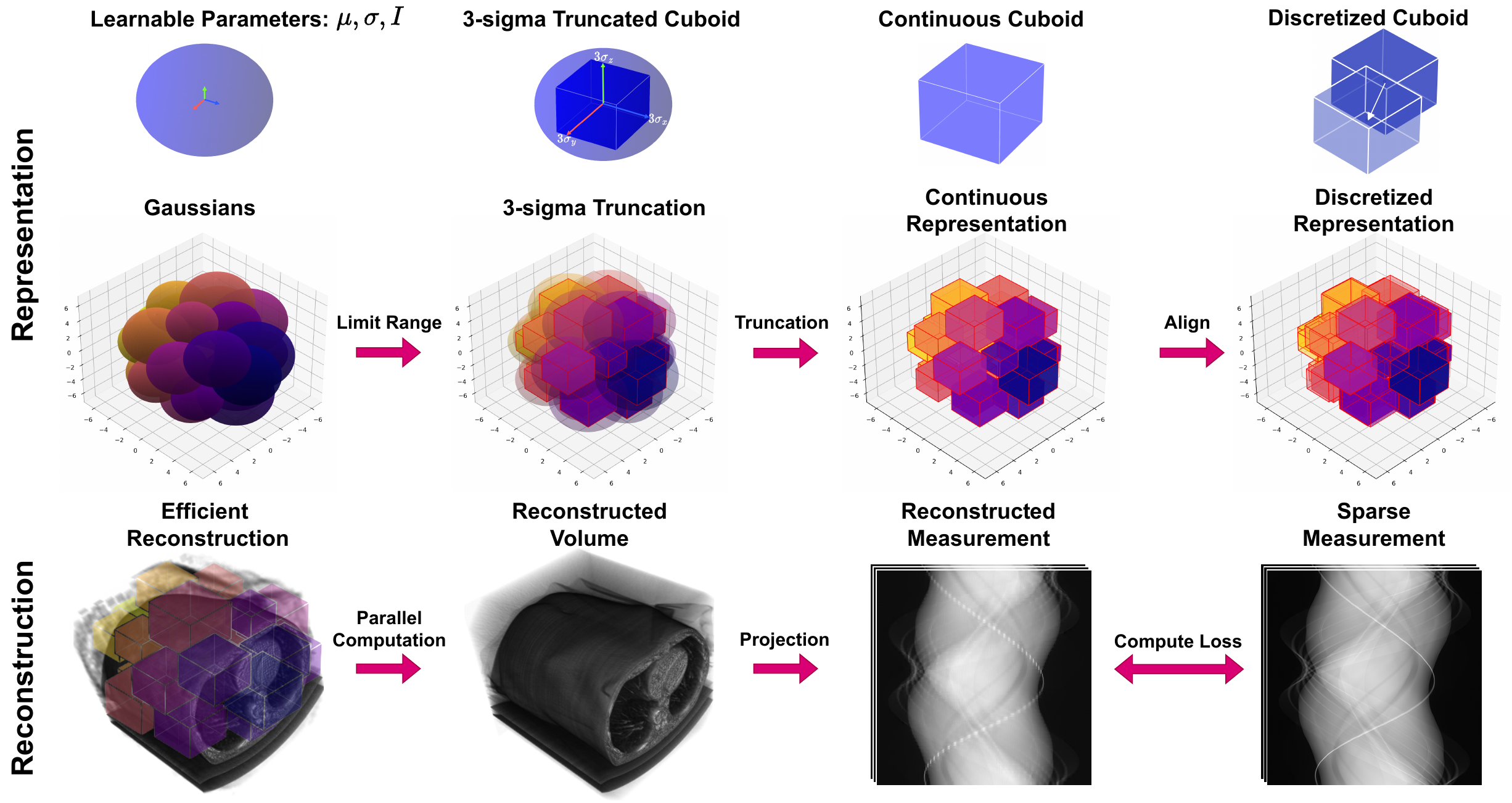}
    \caption{Framework of our baseline solution GIFT.}
    \label{framework}
    \vspace{-3pt}
\end{figure}

Our Gaussian Iteration framework for Tomography (GIFT, Figure~\ref{framework}) is inspired by the development of 3D Gaussian Splatting (3DGS)~\cite{3DGS,car,dif-gaussian,x-gaussian} and is modified from the strong baseline proposed by Wu \textit{et al.}~\cite{DGR}. The core idea is to iteratively reconstruct the 3D volume from a set of Gaussians, transform it into the projection domain, and optimize it within that domain. Since GIFT is intended primarily as a baseline solution, we provide reproducible code and comparative analyses of its innovations on our project page.

\noindent \textbf{Gaussian Representation}
We represent the 3D medical volume as the sum of a set of Gaussian functions. Each Gaussian function is characterized by its center at a mean value $\bm\mu$ and a covariance $\bm\Sigma$ as defined below:
\begin{equation}
  G(\x, \bm{\mu}, \bm\Sigma) = \exp\left(-\frac{1}{2}(\x-\bm\mu)^{\top}\bm\Sigma^{-1}(\x-\bm\mu)\right),
\end{equation}
where $\x \in \mathbb{R}^{d}$ represents a 3D point in the scene, exhibiting a bell-shaped curve symmetrically distributed around the mean $\bm\mu$. The spread of this function in 3D space is determined by the covariance matrix $\Sigma$.

Naively, we can formulate the reconstruction process of $n$ Gaussians as follows:
\begin{equation}
  \mathbf{V} = \sum_{i=1}^{n} G(\x,\bm\mu_i,\bm\Sigma_i)\cdot I_i 
    = \sum_{i=1}^{n} e^{-\frac{1}{2}(\x-\bm\mu_i)^{\top}\bm\Sigma_i^{-1}(\x-\bm\mu_i)}\cdot I_i
    = \sum_{i=1}^{n} e^{-\frac{1}{2}D^2_i}\cdot I_i,
\end{equation}
In this equation, $I_i$ denotes the intensity of the $i$-th Gaussian. This intensity serves dual purposes: it represents the intensity of the voxel in the volume and also acts as the weight of the Gaussian. The term $(\x-\bm\mu)^{\top}\bm\Sigma^{-1}(\x-\bm\mu)$ is recognized as the squared Mahalanobis distance, and we denote it as $D^2_i$ for the $i$-th Gaussian for brevity.

However, this formulation is computationally expensive, as it requires the computation of the squared Mahalanobis distance for each voxel in the volume. To address this issue, we localize the Gaussian mapping to accelerate the reconstruction.

\noindent \textbf{3-sigma Confined Reconstruction} According to the well-known 3-sigma rule, in a Gaussian distribution, the probability of a point being within three standard deviations of the mean is approximately 99.73\%. This implies that the influence of a Gaussian on a voxel decreases with increasing distance from the Gaussian center to the voxel. By considering only the contributions of Gaussians within a specified proximity of each voxel, we can accelerate the reconstruction process.
\label{3sigma}

Specifically, we take the median standard deviation of the Gaussians for all three dimensions as the representative standard deviation $\tilde{\bm\sigma} = (\tilde\sigma_x, \tilde\sigma_y, \tilde\sigma_z)$, and set the size of the neighborhood $\bm\delta$ to be three times this median standard deviation. In formal terms:

Denote the target discretized 3D volume as $\mathbf{V} \in \mathbb{R}^{C \times H \times W}$ where $C$, $H$, and $W$ represent the size of the three dimensions, and denote the neighborhood around the $i$-th Gaussian as $\bm\delta_i \in \mathbb{R}^{c \times h \times w\times d}$ where $c$, $h$, and $w$ represent the size of the neighborhood, and $d=3$ represents the dimension of the 3D coordinates. Note that the neighborhood is centered at the Gaussian center $\bm\mu_i$, thus the distances from the points in $\bm\delta_i$ to the center $\bm\mu_i$ form \textbf{a constant tensor} for all Gaussians since the relative distances from points to the center are fixed. We denote this tensor as $\bm\delta' = \bm\delta_i - \bm\mu_i$ with broadcasting applied, where each point $\bm{p}$ in $\bm\delta_i$ and its corresponding point after transformation $\bm{p}'$ in $\bm\delta'$ satisfies $\bm{p}' = \bm{p} - \bm\mu_i$. 

The computation of $D^2_i$ between the voxel and the Gaussian's mean can hereby be simplified as:
\begin{equation}
    D^2_i = \bm\delta'^{\top} \bm\Sigma_i^{-1} \bm\delta'.
\end{equation}

\begin{figure*}[t]
    \centering
    \setlength{\belowcaptionskip}{-5pt}
    \includegraphics[width=1\linewidth]{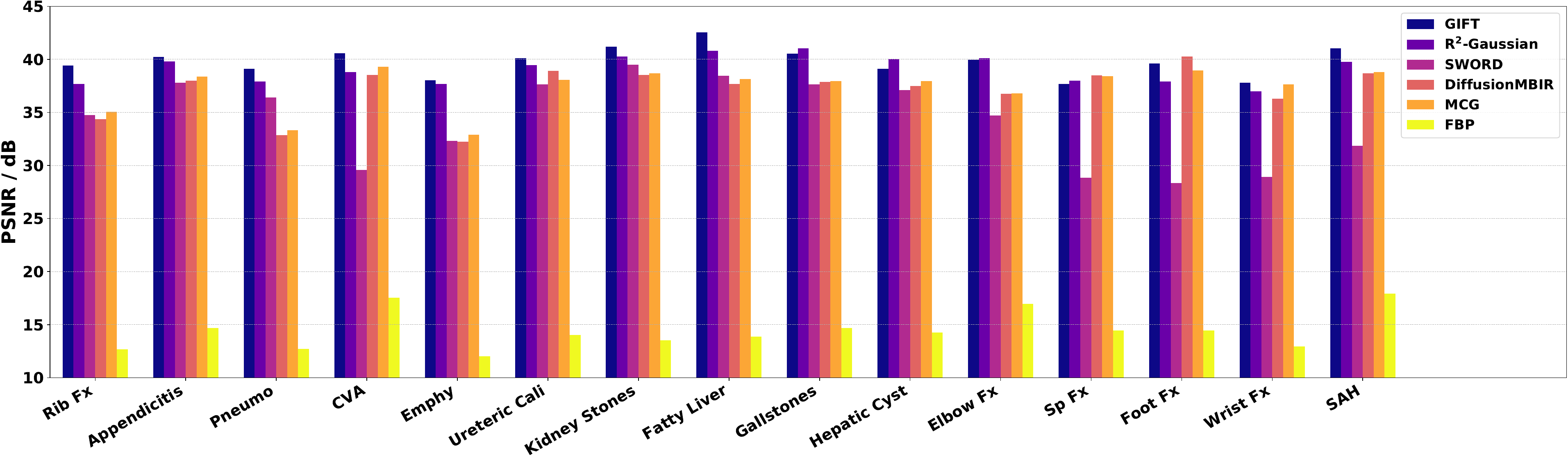}
\caption{Performance of 60-view SV-CT on 15 types of lesions within the \textbf{MORE} dataset.}
\label{column}
\end{figure*}

The computation above does not take the discretized grid into account, which is essential for the reconstruction process. The discretized 3D volume $\mathbf{V}$ is composed of integer coordinates, whereas $\bm\mu_i$ is continuous. Direct discretization of $\bm\mu_i$ to the nearest integer for indexing would render the reconstruction process non-differentiable. To address this, we compute each Gaussian's contribution at the discretized grid instead of its continuous position. We denote $\bm\delta''_i$ as the discretized neighborhood around the Gaussian center. The relationship between $\bm\delta'_i$, $\bm\delta''_i$, and $\bm\mu_i$ is given by:
\begin{equation}
    \bm\delta''_i = \bm\delta'_i - (\bm\mu_i - \left \lfloor \bm\mu_i \right \rfloor) = \bm\delta'_i - \Delta \bm\mu_i,
    \label{eq:alignment}
\end{equation}
where we denote $\Delta \bm\mu_i = \bm\mu_i - \left \lfloor \bm\mu_i \right \rfloor$ for brevity. Each point $\bm{p}$ in $\bm\delta'_i$ and its corresponding point after transformation $\bm{p}'$ in $\bm\delta''_i$ satisfies $\bm{p}' = \bm{p} - (\bm\mu_i - \left \lfloor \bm\mu_i \right \rfloor)$. From now on, we use subscripts to denote tensor dimensions to represent broadcasting operations and tensor-wise operations. For example, Equation \ref{eq:alignment} will be written as $\bm\delta''_{n,c,h,w,d} = \bm\delta'_{c,h,w,d} - \Delta \bm\mu_{n,1,1,1,d}$. Here, $\bm\delta''_{n,c,h,w,d}$ is the tensor comprised of the neighborhoods of all $n$ Gaussians, and $\Delta \bm\mu_{n,1,1,1,d}$ implicitly denotes the expansion of $\Delta \bm\mu_{n,d}$ to identical dimensions for element-wise subtraction.

On the discretized 3D grid, the computation of the squared Mahalanobis distance tensor $D^2_{n,c,h,w}$ can be formulated as an Einstein summation:
\begin{equation}
    D^2_{n,c,h,w} = \sum_{d}^{} \bm\delta{''}_{n,c,h,w,d}^{\top} \bm\Sigma_{n,d,d}^{-1} \bm\delta''_{n,c,h,w,d}
    \label{eq:sum}
\end{equation}

By combining Equations \ref{eq:alignment} and \ref{eq:sum}, we decompose the large Einstein summation above into the sum of four smaller Einstein summations:

\begin{align}
    D^2_{n,c,h,w} = \sum_{d}^{} \bm\delta'_{c,h,w,d} \bm\Sigma^{-1}_{n,d,d} \bm\delta'_{c,h,w,d}  - \sum_{d}^{}  \bm\delta'_{c,h,w,d} \bm\Sigma^{-1}_{n,d,d} \Delta \bm\mu_{n,1,1,d}, \nonumber \\
    -  \sum_{d}^{} \Delta \bm\mu_{n,1,1,d} \bm\Sigma^{-1}_{n,d,d} \bm\delta'_{c,h,w,d} + \sum_{d}^{} \Delta \bm\mu_{n,1,1,d} \bm\Sigma^{-1}_{n,d,d} \ \Delta \bm\mu_{n,1,1,d}.
\end{align}

Then we can compute the contributions of all Gaussians, denoted as $\Gamma_{n,c,h,w}$, as follows:
\begin{equation}
  \Gamma_{n,c,h,w} = e^{-\frac{1}{2}D^2_{n,c,h,w}}\cdot I_{n}
\end{equation}

Note that $\Gamma_{n,c,h,w}$ represents the contributions of all Gaussians within their neighborhoods, and the final step is to add up all the contributions to their corresponding voxels in the volume. A direct way is to loop over each Gaussian and add its contribution to the volume as $\mathbf{V}[\bm\delta_i] \leftarrow  \mathbf{V}[\bm\delta_i] + \Gamma_{i}$. For parallel computation, we can use the scatter-add operation implemented by PyTorch to achieve the same effect in a more efficient way:
\begin{equation}
  \mathbf{V}_{c,h,w} = scatter\_add(\Gamma_{n,c,h,w}, \bm\delta_{n,c,h,w,d}).
\end{equation}

\noindent \textbf{Optimization}
After the 3D volume is reconstructed, we transform the 3D volume to the measurement domain and directly optimize it under the supervision of the current patient's measurement. The transformation $\mathcal{F}$ from the 3D volume to the measurement domain is achieved through the Radon transform.
\begin{equation}
  \hat \M = \mathcal{F}(\mathbf{V}) = Radon(\mathbf{V})
\end{equation}
where $\hat \M$ is the estimated measurement. Then, the optimization problem is to minimize the discrepancy between the estimated measurement $\hat \M$ and the sparse measurement $\M$. We penalize the discrepancy in the measurement domain using the $L_1$ norm and the Structural Similarity Index (SSIM)~\cite{ssim, nilsson2020understanding}. In addition, we add a total variation (TV)~\cite{rudin1992nonlinear, ADMM-TV} regularization term to the 3D volume to preserve its structure. The optimization problem can be formulated as:
\begin{equation}
  \min_{\mathbf{V}} \lambda_1 \left\| \hat \M - \M \right\|_1 + \lambda_2 \text{SSIM}(\hat \M, \M) + \lambda_3 \text{TV}(\mathbf{V}),
\end{equation}

where $\lambda_1=0.4$, $\lambda_2=0.1$, and $\lambda_3=0.5$ are hyperparameters to balance the three terms. Iteratively, we update the parameters of the Gaussians to minimize the objective function. The optimization process is conducted in an end-to-end manner, and the final 3D volume is obtained after convergence.

\begin{table}[t]
  \centering
  \small
   \setlength{\belowcaptionskip}{-5pt}
  \setlength{\abovecaptionskip}{0pt}
  \caption{Average Performance on the MORE Dataset}
  \setlength{\tabcolsep}{0.3mm}
  \resizebox{\linewidth}{!}{
    \begin{tabular}{lccccccccc}
  \toprule
  \multicolumn{1}{c}{\multirow{2}{*}{Method}} & \multirow{2}{*}{Pretrain} & \multicolumn{2}{c}{180-view} & \multicolumn{2}{c}{120-view} & \multicolumn{2}{c}{90-view} & \multicolumn{2}{c}{60-view} \\ 
  \cline{3-10} 
   &  & PSNR & SSIM & PSNR & SSIM & PSNR & SSIM & PSNR & SSIM \\ 
  \hline
   RED-CNN~\cite{REDCNN} & \checkmark  & 33.90 	&0.845 	&32.78 	&0.827 	&30.09 	&0.777 	&30.09 	&0.770 
 \\
  MCG~\cite{MCG} & \checkmark  & 37.29 	&0.883 	&37.44 	&0.884 	&37.53 	&0.884 	&37.39 	&0.883 
 \\
  DiffusionMBIR~\cite{DiffusionMBIR} & \checkmark & 36.84 	&0.958 	&37.49 	&0.958 	&37.36 	&0.958 	&37.13 	&0.957 
 \\
  SWORD~\cite{SWORD} & \checkmark  & 40.30 	&0.930 	&38.74 	&0.917 	&36.92 	&0.902 	&34.25 	&0.872 
 \\
  % \hdashline
  FBP~\cite{fbp1967} & \texttimes & 21.94 	&0.460 	&18.80 	&0.409 	&16.89 	&0.373 	&14.42 	&0.322 
 \\
  NeRP~\cite{NeRP} & \texttimes & 26.87 	&0.803 	&26.74 	&0.800 	&26.86 	&0.805 	&26.66 	&0.799 
 \\
  R$^2$-Gaussian~\cite{r2-gaussian} & \texttimes & 41.14 	&0.967 	&40.44 	&0.962 	&39.75 	&0.958 	&39.08 	&0.954 
 \\
  GIFT (Ours) & \texttimes & \textbf{42.52} 	&\textbf{0.978} 	&\textbf{41.72} 	&\textbf{0.976} 	&\textbf{40.90} 	&\textbf{0.972}	&\textbf{39.80} 	&\textbf{0.968}
 \\
  \bottomrule
  \end{tabular}
  }
  
  \vspace{-5pt}
  \label{tab:average}
  \end{table}

\vspace{-3pt} \section{Experiments}
\label{sec:experiments}
We extensively evaluate various types of methods on the \textbf{MORE} dataset. To ensure the reproducibility of our experiments, we provide the \textbf{code} on our Project Page 

\noindent \textbf{Training Settings} The benchmark setting follows the description in Section~\ref{sec:benchmark}. For training-based methods, we use the training set of 58,528 CT slices to train the model and then evaluate the performance on the test set of 7,498 CT slices. The training set is split into 80\% for training and 20\% for validation. The validation set is used to tune the hyperparameters and select the best model for evaluation. For optimization-based methods, we directly evaluate the performance on the test set without using any training data. Evaluation is conducted on the 15 different lesions in the test set.

\noindent \textbf{Implementation Details} We follow the original implementation of each method and use the default hyperparameters as specified in our code. We implement our GIFT framework in PyTorch and use the Adam optimizer with a learning rate of 3e-4. The training is conducted on the sparse measurements of the current patient directly. We set the number of Gaussians to 150K at initialization and optimize them iteratively.  For all experiments, we use an NVIDIA RTX 6000 Ada Generation GPU with 48GiB memory.

\noindent \textbf{Benchmark Results on MORE}

In Table~\ref{tab:average}, we present the average performance across 15 different lesion types on the MORE dataset. The results are reported in terms of Peak Signal-to-Noise Ratio (PSNR)~\cite{psnr} and Structural Similarity Index (SSIM)~\cite{ssim} for the 180-view, 120-view, 90-view, and 60-view tasks. Since the detailed data corresponds to 15 Tables, \textbf{\textit{we include all these Tables in our Project Page}} . In Figure~\ref{column}, we show the PSNR of 60-view sparse-view CT on all lesions. Notably, our GIFT achieves superior performance across most lesion types.

It can be observed that even without using pretrained data, R$^2$-Gaussian and our GIFT achieve advanced performance compared to those with training data. 
Our benchmark demonstrates the potential of optimization-based methods in achieving strong generalization ability and robustness to unseen anatomies and lesions. 

\begin{table}[t]
  \centering
  \small
  \setlength{\belowcaptionskip}{0pt}
  \setlength{\abovecaptionskip}{3pt}
  \caption{Generalization Comparison on AAPM-Mayo (Abdomen) trained with different datasets. * denotes trained on AAPM-Mayo dataset, and the rest are trained on MORE.}
  \setlength{\tabcolsep}{0.6mm}
  \begin{tabular}{lccccccccc}
  \toprule
  \multicolumn{1}{c}{\multirow{2}{*}{Method}} & \multicolumn{2}{c}{180-view} & \multicolumn{2}{c}{120-view} & \multicolumn{2}{c}{90-view} & \multicolumn{2}{c}{60-view} \\ 
  \cline{2-9} & PSNR & SSIM & PSNR & SSIM & PSNR & SSIM & PSNR & SSIM \\ 
  \hline
   RED-CNN*~\cite{REDCNN}& 40.08 & 0.976 & 37.42 & 0.973 & 36.27 & 0.965 & 34.88 & 0.957 \\
  MCG*~\cite{MCG} & 40.42 & 0.969 & 39.57 & 0.960 & 38.02 & 0.935 & 37.17 & 0.921 \\
  DiffusionMBIR*~\cite{DiffusionMBIR} & 41.78 & 0.990 & 40.83 & 0.964 & 39.98 & 0.942 & \textbf{38.67} & 0.932 \\
  SWORD*~\cite{SWORD}  & 45.08 & 0.994 & 42.49 & 0.990 & 41.27 & 0.986 & 38.49 & \textbf{0.978} \\
  \hdashline
   RED-CNN~\cite{REDCNN}& 41.73 & 0.978 & 39.55 & 0.981 & 38.67 & 0.978 & 36.70 & 0.971 \\
  MCG~\cite{MCG} & 41.13 & 0.975 & 40.38 & 0.971 & 38.49 & 0.949 & 37.36 & 0.919 \\
  DiffusionMBIR~\cite{DiffusionMBIR} & 42.57 & 0.991 & 41.22 & 0.968 & 40.05 & 0.944 & 38.82 & 0.934 \\
  SWORD~\cite{SWORD}  & \textbf{46.03} & \textbf{0.995} & \textbf{44.60} & \textbf{0.992} & \textbf{41.58} & \textbf{0.988} & 38.42 & 0.976 \\
  \bottomrule
  \end{tabular}
  \vspace{-10pt}
  \label{tab:svct-aapm}
  \end{table}

\noindent \textbf{Generalization Ability Evaluation}  

To further demonstrate the strength of our MORE dataset, we conduct a generalization ability comparison on two different datasets: the MORE dataset and the AAPM-Mayo dataset~\cite{mayo}, which is a widely used dataset for sparse-view CT reconstruction~\cite{PLA, SWORD, DiffusionMBIR}. We conduct the two groups of experiments described below:

\noindent \textbf{Trained on different datasets, Test on AAPM-Mayo} In this group of experiments, we train the learning-based methods on the AAPM-Mayo dataset and our MORE dataset, and then evaluate the performance on the abdomen part of the AAPM-Mayo dataset. The results are shown in Table~\ref{tab:svct-aapm}.  
It can be observed that the methods trained on our MORE dataset generally achieve better performance than those trained on the AAPM-Mayo dataset. It is important to note that the AAPM-Mayo training and test sets are from the same distribution, while our MORE dataset represents a different distribution.
Despite this, the methods trained on our MORE dataset still outperform those trained on the AAPM-Mayo dataset, demonstrating that our MORE dataset can effectively improve the generalization ability of deep learning models.

\noindent \textbf{Trained on different datasets, Test on MORE} In this group of experiments, we train the learning-based methods on the AAPM-Mayo dataset and our MORE dataset, and then evaluate the performance on the foot fracture part of the MORE dataset. The results are shown in Table~\ref{tab:svct-foot-fracture}. This time, the methods trained on our MORE dataset achieve greater advantages than those trained on the AAPM-Mayo dataset. This is because our MORE dataset contains a wider range of anatomies and lesions, which can help the methods to learn more robust features.

\begin{table}[t]
  \centering
  \small
  \setlength{\belowcaptionskip}{0pt}
  \setlength{\abovecaptionskip}{3pt}
  \caption{Generalization Comparison on MORE (Foot Fracture) trained on different datasets. * denotes trained on the AAPM-Mayo dataset, and the rest are trained on MORE.}
  \setlength{\tabcolsep}{0.6mm}
  \begin{tabular}{lccccccccc}
  \toprule
  \multicolumn{1}{c}{\multirow{2}{*}{Method}} & \multicolumn{2}{c}{180-view} & \multicolumn{2}{c}{120-view} & \multicolumn{2}{c}{90-view} & \multicolumn{2}{c}{60-view} \\ 
  \cline{2-9} & PSNR & SSIM & PSNR & SSIM & PSNR & SSIM & PSNR & SSIM \\ 
  \hline
   RED-CNN*~\cite{REDCNN}& 29.57 & 0.717 & 29.10 & 0.684 & 28.01 & 0.627 & 26.83 & 0.591 \\
  MCG*~\cite{MCG} & 33.56 & 0.823 & 33.41 & 0.817 & 33.18 & 0.795 & 32.76 & 0.788 \\
  DiffusionMBIR*~\cite{DiffusionMBIR} & 35.48 & 0.899 & 35.46 & 0.895 & 35.47 & 0.896 & 35.45& 0.893 \\
  SWORD*~\cite{SWORD}  & 30.01 & 0.764 & 29.63 & 0.758 & 28.22 & 0.729 & 26.58 & 0.633 \\
  \hdashline
   RED-CNN~\cite{REDCNN}  & 37.53 & 0.860 & 35.61 & 0.783 & 32.52 & 0.817 & 32.46 & 0.837 \\
  MCG~\cite{MCG} & 39.40 & 0.891 & 39.62 & 0.895 & 39.43 & 0.894 & 39.45 & 0.894 \\
  DiffusionMBIR~\cite{DiffusionMBIR}& \textbf{40.45} & \textbf{0.956} & \textbf{40.31} & \textbf{0.955} & \textbf{40.22} & \textbf{0.954} & \textbf{40.26}& \textbf{0.957} \\
  SWORD~\cite{SWORD}  & 34.92 & 0.927 & 36.40 & 0.905 & 31.95 & 0.866 & 28.33 & 0.783 \\
  \bottomrule
  \end{tabular}
  \label{tab:svct-foot-fracture}
  \vspace{-10pt}
  \end{table}
\section{Conclusion}
In this paper, we propose the MORE dataset, a comprehensive collection of CT scans for medical image reconstruction research. The dataset includes a diverse range of scans covering various anatomies and conditions, making it a valuable resource for developing and testing reconstruction algorithms. We also introduce a baseline solution, GIFT, for 3D volume reconstruction, which leverages the efficiency of localized Gaussian mapping to accelerate the reconstruction process. Our extensive benchmark shows stronger generalization ability over previous datasets.

\begin{acks}
This work is supported by the Fundamental Research Funds for the Central Universities (project number YG2024ZD06), National Natural Science Foundation of China (No. 62176155), and Shanghai Municipal Science and Technology Major Project (2021SHZDZX0102).
\end{acks}

\bibliographystyle{ACM-Reference-Format}
\balance
\bibliography{main}

\end{document}